\documentclass[aps,pre,preprint,superscriptaddress]{revtex4-1}
\usepackage{graphicx,amsmath,amsfonts,amssymb}

\begin{document}

\title{Quantum phase transition in an effective three-mode model of 
interacting bosons}

\author{H. M. Fraz\~ao}
\email{heliques@ufpi.edu.br}
\affiliation{Universidade Federal do Piau\'{\i}, Campus Profa. Cinobelina
Elvas, Bom Jesus, PI, Brazil}
\affiliation{Departamento de F\'{\i}sica, Instituto de Ci\^encias Exatas, 
Universidade Federal de Minas Gerais, Belo Horizonte, MG, Brazil}

\author{J. G. Peixoto de Faria}
\affiliation{Departamento de Matem\'{a}tica, Centro Federal
de Educa\c{c}\~{a}o Tecnol\'{o}gica de Minas Gerais, 
Belo Horizonte, MG, Brazil.}

\author{G. Q. Pellegrino}
\affiliation{Departamento de Matem\'{a}tica, Centro Federal
de Educa\c{c}\~{a}o Tecnol\'{o}gica de Minas Gerais, 
Belo Horizonte, MG, Brazil.}

\author{M. C. Nemes}
\affiliation{Departamento de F\'{\i}sica, Instituto de Ci\^encias Exatas, 
Universidade Federal de Minas Gerais, Belo Horizonte, MG, Brazil}

\date{\today}

\begin{abstract}

In this work we study an effective three-mode model describing interacting 
bosons. These bosons can be considered as exciton-polaritons in a semiconductor 
microcavity at the magic angle. This model exhibits quantum phase transition 
(QPT) when the parameters of the corresponding Hamiltonian are continuously 
varied. The properties of the Hamiltonian spectrum ({\it e.g.}, the distance 
between two adjacent energy levels) and the phase space structure of the 
thermodynamic limit of the model are used to indicate QPT. The relation between 
spectral properties of the Hamiltonian and the corresponding classical frame of 
the thermodynamic limit of the model is established as indicative of QPT . The 
average number of bosons in a specific mode and the entanglement properties of 
the ground state as functions of the parameters are used to characterize the 
order of the transition and also to construct a phase diagram. Finally, we 
verify our results for experimental data obtained for a setting of 
exciton-polaritons in a semiconductor microcavity.

\end{abstract}

\maketitle

\section{Introduction}

One issue of great interest in condensed matter physics today is quantum phase 
transition (QPT) \cite{QPT}. Differently from the usual thermodynamic phase 
transition, which is guided by thermodynamic fluctuations and characterized by 
a critical temperature, a QPT can be observed for $T=0$ and is conducted by 
quantum fluctuations. The change in the system due to a QPT is observed when 
some parameter of the Hamiltonian is varied, rather than temperature as in the 
thermodynamic phase transition. As a simple example, we have the quantum Ising 
model in a transverse field, where a QPT is observed between ferromagnetic and 
paramagnetic phases when the intensity of the applied field is varied 
\cite{QPT,ising}. As this example illustrates, the study of QPT conveys a 
better 
understanding of the complex behavior shown by many-body systems. This study 
can 
cross very different systems as, for example, systems involving light-matter 
interactions, such as cavity arrays coupled by optical fibers \cite{arda}, or a 
two-species condensate of interacting bosons trapped in optical lattices 
\cite{lingua}. The next paragraphs present other instances where QPT in 
many-body problems is considered and which are of interest for this work.

One of the systems under intense attention in recent decades is the 
semiconductor microcavity \cite{kavokin}. In this system, the interaction 
between cavity photons and excitons belonging to the semiconductor gives rise 
to 
a new {\em quasi}-particle called exciton-polariton \cite{kavokin2}. There are 
many interesting features shown by this system as, for example, the 
superfluidity \cite{sfluid}, and the generation of a Bose-Einstein condensate 
(BEC) in a solid state system \cite{bec} (which occurs even at room temperature 
\cite{bec2}). Another interesting feature is the so-called {\it magic angle} 
configuration in which we observe a parametric amplification of the emitted 
light \cite{savidis}. One possible theoretical description of this feature is 
given by considering only three modes for the exciton-polaritons, denominated 
{\it pump, signal} and {\it idler} \cite{ciuti}. The evidences pointed by 
experiment \cite{savidis} suggest that only these three modes are coherently 
and 
macroscopicaly populated \cite{kavokin,ciuti}. In this situation one can 
approximate the Hamiltonian of the system by an effective three-mode 
Hamiltonian 
as we do in this work.

The experimental observation of QPT in the scope of condensed matter physics 
has been boosted by the development of techniques that allowed for the 
storage and handling of matter at the atomic level. In the past two decades, 
intense research has been done on systems of interacting trapped bosons, 
especially cold atoms trapped in optical lattices \cite{opt_latt}. The interest 
in such systems is assigned to the their capacity to simulate many phenomena 
predicted to occur in arrangements of trapped atoms, such as the BEC itself 
\cite{bec3}, solitons \cite{solitons}, bosonic Josephson effect, and nonlinear 
oscillations \cite{nlinear}, and also a QPT from a Mott-insulator-like phase to 
a superfluid-like one \cite{QPT,ott}. For a two-well condensate \cite{dwell} or 
a three-well condensate \cite{twell,viscondi}, the transition is between two 
dynamical regimes: macroscopic self-trapping and Rabi (or Josephson) 
oscillations 
\cite{rmp73}. These kinds of systems are generally described by a 
Bose-Hubbard-type Hamiltonian \cite{hubbard,paredes,referee}. In this work 
we study an effective three-mode Hamiltonian essentially different from 
a Bose-Hubbard Hamiltonian but which exhibits a QPT between two phases: the 
macroscopic self-trapping (MST), characterized by the vanishing of the 
tunneling 
between the modes, and {a regime of oscillations} (RO), where tunneling is 
present.

In recent years, also classical analyses involving the thermodynamic limit of 
some quantum many-body models have contributed to the investigations of QPT. 
These analyses come to add themselves to other valuable tools to investigate 
QPT, many of them referring to the properties of the spectrum of the 
corresponding quantum Hamiltonian, as the level approximation (crossing), or to 
the measurement of entanglement of the ground state \cite{entanglement} near 
the 
critical point. Different systems were studied in this quantum-classical 
context, {\it e.g.}, the Lipkin model \cite{lipkin} and the pairing model 
\cite{reis} (both in nuclear physics), the Dicke model for superradiance in 
quantum optics \cite{dicke1}, excitons in semiconductor bilayer electron 
systems 
\cite{gian}, ultracold Bose gases trapped in multiple wells \cite{viscondi}, 
among others \cite{hines}. In parallel with the quantum treatment, we perform 
as 
well the classical limit analysis to study QPT in the effective three-mode 
model 
for a system of interacting bosons. Our analysis results in a close analogy 
between some aspects of the spectrum of the quantum Hamiltonian and classical 
properties of the corresponding thermodynamic limit of the model. As an 
example, 
the different phases in the quantum regime are related to the existence of 
closed orbits in the classical thermodynamic limit. In fact, the MST phase 
corresponds to the situation in which there are no closed orbits in phase 
space, 
whereas the RO phase is characterized by the presence of such orbits. In this 
way, separatrices delimiting regions of the phase space characterized by 
different dynamical regimes in the classical limit are identified with the 
presence of non-trivial minima in the separation of adjacent energy levels. 
Besides that, the properties of the ground state are used to characterize the 
order of the transition and also to construct a phase diagram that shares some 
similarity with the phase diagram obtained for the Lipkin model \cite{lipkin}. 
Our results can be applied to describe a system of exciton-polaritons 
in a semiconductor microcavity at the magic angle.
 
This work is organized as follows. In Sec. II, the effective Hamiltonian is 
presented and, in Sec. III, its spectral properties are shown; in Sec. 
IV, the classical thermodynamic limit of the Hamiltonian is taken and the 
resulting phase space is analyzed. In Sec. V, we discuss the QPT and a phase 
diagram. In Sec. VI, we verify our results for exciton-polaritons in a 
semiconductor microcavity, based on available experimental data. Finally, we 
present our conclusions in Sec. VII.

\section{Three-mode approximation for interacting bosons}

The starting point of this work is a three-mode Hamiltonian for interacting 
bosons in the form \begin{equation} \label{H0} H=\sum_{i=0,1,2}E_ia_i^{\dag}a_i
+\sum_{i+j=k+l}\hbar G_{ijkl}a_i^{\dag}a_j^{\dag}a_ka_l,\end{equation} where 
$a_{i}^{\dag}\;(a_{i})$ is the creation (annihilation) operator for a boson 
with 
energy $E_{i}$. The first term describes the free bosons while the second one 
describes the interaction between different bosonic modes which obeys the 
condition: $G_{ijkl}\neq0$ for $i+j=k+l$, $G_{ijkl}=0$ for $i+j\neq k+l$. As we 
see in the following, we can obtain a Hamiltonian as Eq. \eqref{H0} for the 
description of exciton-polaritons in a semiconductor microcavity at the magic 
angle (Sec. II.A). An effective Hamiltonian (Sec. II.B) can be obtained 
from Eq. \eqref{H0} by using the conservation of total number of bosons and the 
condition $i+j=k+l$ which means conservation of momentum for exciton-polaritons.

\subsection{Exciton-polaritons in a semiconductor microcavity at the 
magic angle}

An exciton-polariton is a {\it quasi}-particle formed from the coupling between 
an exciton and a photon in a semiconductor microcavity. Under certain 
experimental conditions we can model the exciton-polaritons in a semiconductor 
microcavity by the Hamiltonian \cite{kavokin,ciuti} 
\begin{equation}\label{Hlow} 
H=\sum_{\mathbf{k}}\hbar\Omega_{\mathbf{k}}p^{\dag}_{\mathbf{k}}p_{\mathbf{k}}+ 
\frac{1}{2}\sum_{\mathbf{k,k',q}}V^{PP}_{\mathbf{k,k',q}} 
p^{\dag}_{\mathbf{k+q}}p^{\dag}_{\mathbf{k'-q}}p_{\mathbf{k}}p_{\mathbf{k'}}, 
\end{equation} where $p_{\mathbf{k}}^{\dag}\;(p_{\mathbf{k}})$ is the creation 
(annihilation) operator for an exciton-polariton with in-plane wave-vector 
$\mathbf{k}$ and energy $\hbar\Omega_{\mathbf{k}}$. The first term describes 
the 
free exciton-polaritons while the second one describes the interaction between 
different exciton-polariton modes. For typical values of the experimental 
parameters, the interaction coefficients are given by 
\begin{equation}\label{Vpp2} V^{PP}_{\textbf{k},\textbf{k'},\textbf{q}}\simeq 
V_0 u_{|\textbf{k'-q}|}u_k. u_{|\textbf{k+q}|}u_{k'}, \;V_0=\frac{6e^2 
a_{\mathrm{exc}}}{\epsilon_0 A},\end{equation} where $a_{\mathrm{exc}}$ is the two-dimensional 
Bohr radius of the exciton-polariton, $\epsilon_0$ is the dieletric constant of 
the semiconductor and $A$ is the macroscopic quantization area. The $u_k$'s are 
the so-called Hopfield coefficients, given by\begin{equation}
u_k=\left(\frac{\Delta_k+\sqrt{\Delta_k^2+\Omega_R^2}}{2\sqrt{\Delta_k^2 
+\Omega_R^2}}\right)^{1/2},\end{equation} where 
$\Delta_k=E_{\mathrm{cav}}(k)-E_{\mathrm{exc}}(k)$ 
is the detuning between the energies of cavity photons and excitons.

At the magic angle configuration, the system exhibits a parametric 
amplification 
of the emitted light \cite{savidis}. In this situation, a theoretical 
description of the system is given by considering that only three modes for 
exciton-polaritons ---namely, signal ($\mathbf{0}$), pump ($\mathbf{k}_p$), 
and 
idler ($2\mathbf{k}_p$) ---are coherently and macroscopicaly populated 
\cite{ciuti}. If $\mathbf{k}_p$ is the wave vector of pumping, the scattering 
of 
two $\mathbf{k}_p$ exciton-polaritons results in two other exciton-polaritons 
with wave vectors $\mathbf{0}$ and $2\mathbf{k}_p$. Considering this dynamics, 
we can approximate Hamiltonian Eq. \eqref{Hlow} by a three-mode Hamiltonian as 
Eq. \eqref{H0} with $p_{\mathbf{0}}\equiv a_0$, $p_{\mathbf{k}_p}\equiv a_1$, and 
$p_{2\mathbf{k}_p}\equiv a_2$. In this way, the coefficients $G_{ijkl}$ are 
functions of the $V^{PP}_{\textbf{k},\textbf{k'},\textbf{q}}$ and the condition 
$i+j=k+l$ in Eq. \eqref{H0} is assured by the conservation of momentum in the 
second 
term of Eq. \eqref{Hlow}.

\subsection{An effective Hamiltonian}

The three-mode Hamiltonian Eq. \eqref{H0} conserves the total number of bosons 
represented by the observable $\hat{N}\equiv a_0^{\dag}a_0+a_1^{\dag}a_1+ 
a_2^{\dag}a_2$. Besides that, the difference between the population of bosons 
in 
modes $0$ and $2$, i.e., the imbalance represented by $\hat{D}\equiv 
a_0^{\dag}a_0-a_2^{\dag}a_2$, is also conserved. These are two constants of 
motion under the evolution given by Eq. \eqref{H0}. In terms of $\hat{N}$ and 
$\hat{D}$ operators, we can rewrite Hamiltonian Eq. \eqref{H0} as 
\begin{equation}\label{H1} 
H=H_{ND}(\hat{N},\hat{D})+H_{eff}(\hat{n}_0,a_0a_2a_1^{\dag2}),\end{equation} 
where $\hat{n}_0\equiv a_0^{\dag}a_0$. For convenience and symmetry, we will 
consider $N$ even and $D=0$. We can observe that $H_{ND}$ is constant, giving 
rise only to a global phase in the state of the system as a function of time. 
The time evolution depends only on the second part of Eq. \eqref{H1}, an effective 
Hamiltonian given by \begin{equation}\label{H2} H_{\mathrm{eff}}=\hbar\delta 
\hat{n}_0+\hbar g(\hat{n}_0)^2 
+\hbar[Ga_0^{\dag}a_2^{\dag}a_1^2+G^{*}a_0a_2(a_1^{\dag})^2], \end{equation} 
where $\hbar\delta=E_0+E_2-2E_1-\hbar G_{0000}-\hbar G_{2222} +2\hbar G_{1111}
+4\hbar N[G_{1212}+G_{0101}-G_{1111}]$, 
$g=-8G_{1212}+G_{0000}+4G_{0202}-8G_{0101}+4G_{1111}+ G_{2222}$, and 
$G=2G_{1102}$. In the next section we study the properties of the spectrum of 
the above Hamiltonian for different values of parameters $G/\delta$ and 
$g/\delta$.

\section{Quantum spectrum}

\begin{figure}\centering
  \includegraphics[width=8.5cm,angle=0]{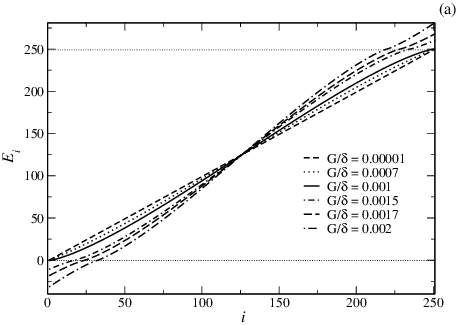}\vspace{.55cm}
  \includegraphics[width=8.5cm,angle=0]{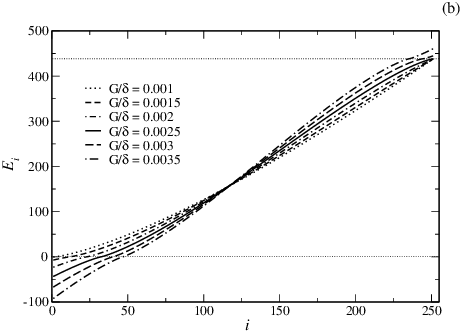}
  \caption{Energy eigenvalues $E_i$ as a function of the eigenstate index $i$ 
for $g/\delta=0$ (a), $g/\delta=0.003$ (b), $N=500$, and different   
values of the three-mode coupling coefficient $G/\delta$. The horizontal   
dotted lines mark the values $E=0$ and $E=N/2+(g/\delta)(N/2)^2$.}
  \label{espectro}
\end{figure}

\begin{figure}\centering
  \includegraphics[width=8.5cm,angle=0]{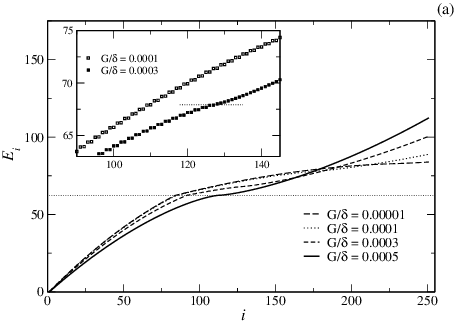}\vspace{.55cm}
  \includegraphics[width=8.5cm,angle=0]{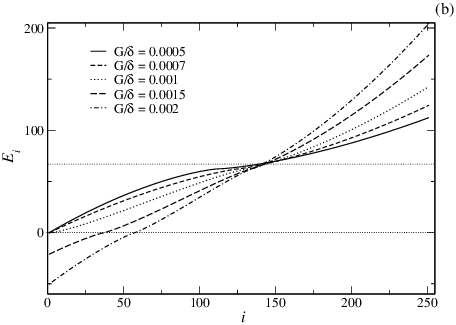}
  \caption{Energy eigenvalues $E_i$ as a function of the eigenstate index $i$ 
for $g/\delta=-0.003$,  $N=500$, and different values of the three-mode 
coupling coefficient $G/\delta$. For $G/\delta<0.0005$ (a) we observe a 
two-fold degeneracy which vanishes for $G/\delta>0.0005$ 
(b).}\label{espectro2}
\end{figure}

In Fig. \ref{espectro} we observe the eigenvalues $E_i$ ($i=1,2,...,N/2+1$) for 
the rescaled Hamiltonian $E=H_{\mathrm{eff}}/\hbar\delta$ as a function of the 
eigenstate 
index $i$ for different values $G/\delta$ of the coupling between the three 
modes. We take $g/\delta=0$ [Fig. \ref{espectro}(a)], $g/\delta=0.003$ 
[Fig. \ref{espectro} (b)], and $g/\delta=-0.003$ [Fig. \ref{espectro2}], 
and we choose $N=500$. In both cases of Fig. \ref{espectro}, the spectra show 
maximal level densities at energies $E = 0$ and $E = N/2 + \left( g/ 
\delta\right)\left( N/2 \right)^2$, which are inflection points. Moreover, we 
see that for $g/\delta=0$ the spectra are always symmetric with respect to the 
central level and both inflection points move accordingly toward the center as 
a function of $G/\delta$. For values $g/\delta > 0$, the spectra are asymmetric 
and the inflection points move separately toward the center as $G/\delta$ is 
varied. For $g/\delta < 0$ [Fig. \ref{espectro2}], we observe a more complex 
behavior with a two-fold degeneracy which disappears for $G/\delta>0.0005$ 
[Fig. 
\ref{espectro2} (a)], but in essence it is similar to the case $g/\delta > 0$, 
with two asymmetrical inflection points moving separately to the centre of the 
spectrum (Fig. \ref{espectro2} (b)). When we calculate the mean values of 
population in mode $0$, $\langle n_0\rangle$, and in mode $1$, $\langle 
n_1\rangle$, for a degenerate spectrum, we find that, at a degeneracy, 
there is the possibility of population inversion between modes with the same 
energy value. In other words, two states with the same energy, $E_i = E_{i + 
1}$, may have $\langle n_0\rangle > \langle n_1\rangle$ or $\langle n_0\rangle 
< \langle n_1\rangle$. Therefore, the degeneracy is characterized by the 
possibility of reversing the population of modes without altering the state's 
energy.

\begin{figure} \centering
  \includegraphics[width=8.5cm,angle=0]{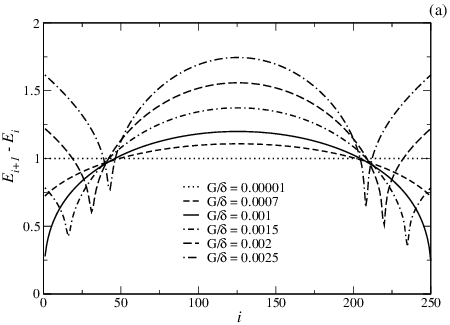}\vspace{.55cm}
  \includegraphics[width=8.5cm,angle=0]{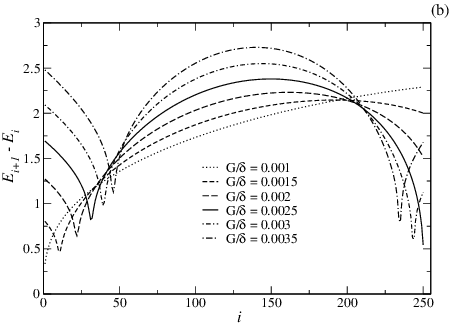}\vspace{.55cm}
  \includegraphics[width=8.5cm,angle=0]{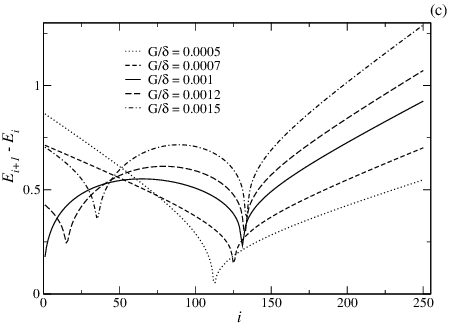}
  \caption{Difference between adjacent energy eigenvalues $E_{i+1}-E_i$ as a   
function of the index $i$ for $g/\delta=0$ (a), $g/\delta=0.003$ (b), 
and $g/\delta=-0.003$ (c), $N=500$, and different values of the  
three-mode coupling coefficient $G/\delta$.}\label{d_espectro}
\end{figure}

We can also observe the behavior of the spectrum in Fig. \ref{d_espectro}, 
where maximal level densities appear as minimal differences for adjacent energy 
levels. It has been shown \cite{reis,gian} that in these Curie-Weiss models a 
quantum phase transition is connected with the level approximation, occurring 
maximally at these inflection points. We will also see that in the 
thermodynamic 
limit the inflection point is associated with a separatrix orbit in the 
corresponding classical phase space.


\section{Classical Thermodynamic Limit of The Hamiltonian}

The classical analog can be obtained as the thermodynamic limit of 
Hamiltonian Eq.
\eqref{H2}. The first step is the mapping of the operators into a SU(2) algebra 
by taking \begin{equation} J_z\equiv \hat{n}_0-\frac{\hat{N}}{4},\;\;\mathrm{ 
and }\end{equation}\begin{equation} 
J_{+}=J_{-}^{\dag}=\frac{1}{\sqrt{2\hat{n}_0(N-2\hat{n}_0+1)}}a_0^{\dag} 
a_2^{\dag}a_1^2.\end{equation} With these operators, Hamiltonian Eq. \eqref{H2} is 
rewritten as \begin{equation}\nonumber H_{\mathrm{eff}}=\hbar\delta(J_z+J)+\hbar 
g(J_z+J)^2 \end{equation} \begin{equation} +\hbar \left\{ 
G\sqrt{2(J_z+J)[N-2(J_z+J)+1]}J_{+} \right. \end{equation}
\begin{equation}\nonumber \left. +G^{*}J_{-}\sqrt{2(J_z+J)[N-2(J_z+J)+1]} 
\right\} \end{equation} with $J=N/4$.

The following step to the thermodynamic limit is obtained by rescaling the 
Hamiltonian by the density $N/V$ and taking the limits $N\rightarrow\infty$ and 
$V\rightarrow\infty$, with the ratio $N/V$ being kept constant. In this limit, 
the classical variables are provided by the usual definitions \cite{lieb} 
\begin{equation} j_k=\lim_{J\rightarrow\infty}\frac{J_k}{J},\;(k=+,-,z) 
\end{equation} and \begin{equation}j_x=\frac{1}{2}(j_{+}+j_{-})=\sqrt{1-j_z^2} 
\cos\phi,\end{equation} where $\phi$ and $j_z$ correspond to canonical 
conjugate 
variables. The classical Hamiltonian finally obtained is written as 
\begin{equation} h(j_z,\phi)=\delta'(j_z+1)+g'(j_z+1)^2 +4 G'(1-j_z^2) 
\cos(\phi) \end{equation} with the rescaled parameters $\delta'=\hbar\delta 
V/4$, $g'=\hbar gNV/16$ and $G'=\hbar GNV/16$.

\begin{figure}\centering
  \includegraphics[width=8.5cm,angle=-0]{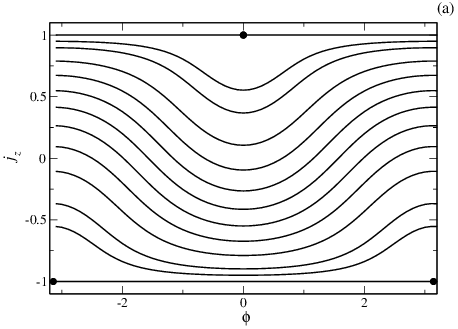}\vspace{.55cm}
  \includegraphics[width=8.5cm,angle=-0]{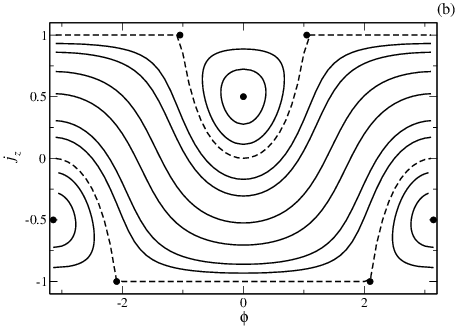}
  \caption{Phase space $\phi \times j_z$ for $g'/\delta'=0$, $G'/\delta'=0.125$
 (a), and $G'/\delta'=0.25$ (b). Rescaled parameters calculated for $N=500$,   
$g/\delta=0$, $G/\delta=0.001$ (a), and $G/\delta=0.002$   
(b). The critical points of $h(\phi,j_z)$ are signaled by black dots.}
  \label{hclass1}
\end{figure}
\begin{figure}\centering
  \includegraphics[width=8.5cm,angle=-0]{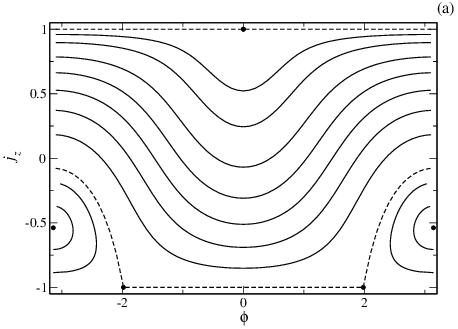}\vspace{.55cm}
  \includegraphics[width=8.5cm,angle=-0]{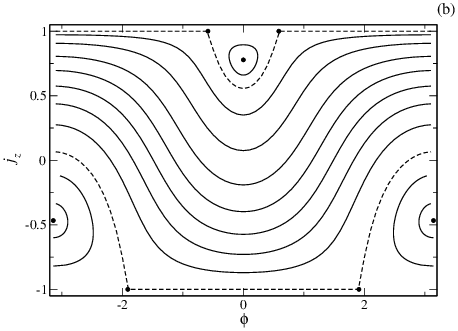}
  \caption{Phase space $\phi \times j_z$ for $g'/\delta'=0.375$, $G'/\delta'=   
0.3125$ (a), and $G'/\delta'=0.375$ (b). Rescaled parameters calculated for   
$N=500$, $g/\delta=0.003$, $G/\delta=0.0025$ (a), and   
$G/\delta=0.003$ (b). The critical points of $h(\phi,j_z)$ are   
signaled by black dots.}\label{hclass2}
\end{figure}

\begin{figure}\centering
  \includegraphics[width=8.5cm,angle=-0]{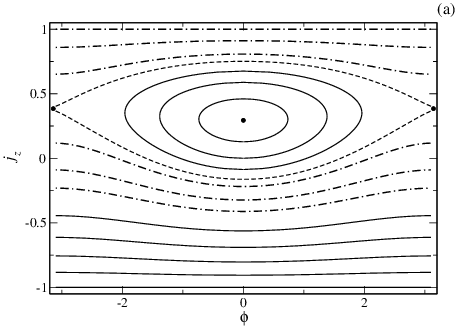}\vspace{.55cm}
  \includegraphics[width=8.5cm,angle=-0]{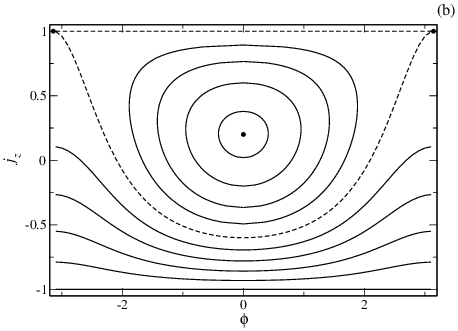}
  \caption{Phase space $\phi \times j_z$ for $g'/\delta'=-0.375$, $G'/\delta'=  
0.0125$ (a), and $G'/\delta'=0.0625$ (b). Rescaled parameters calculated for   
$N=500$, $g/\delta=-0.003$, $G/\delta=0.0004$ (a), and   
$G/\delta=0.0005$ (b). The critical points of $h(\phi,j_z)$ are   
signaled by black dots and the degenerated open orbits by trace-dotted 
curves.}\label{hclass3}
\end{figure}
\begin{figure}\centering
  \includegraphics[width=8.5cm,angle=-0]{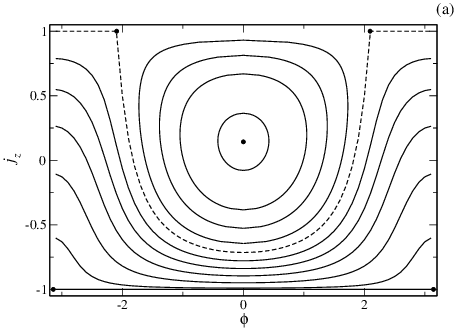}\vspace{.55cm}
  \includegraphics[width=8.5cm,angle=-0]{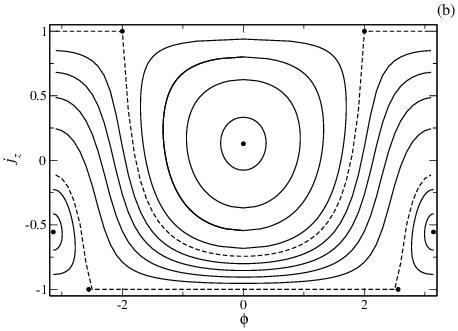}
  \caption{Phase space $\phi \times j_z$ for $g'/\delta'=-0.375$, $G'/\delta'=  
0.125$ (a), and $G'/\delta'=0.15$ (b). Rescaled parameters calculated for   
$N=500$, $g/\delta=-0.003$, $G/\delta=0.001$ (a), and   
$G/\delta=0.002$ (b). The critical points of $h(\phi,j_z)$ are   
signaled by black dots.}\label{hclass4}
\end{figure}

The classical phase space $\phi \times j_z$ is shown in Figs. \ref{hclass1}--\ref{hclass4}, 
using, respectively, $g'/\delta'=0$, $g'/\delta'=0.375$, and 
$g'/\delta'=-0.375$ for some given values of the three-mode coupling 
$G'/\delta'$. As expected, the phase space is periodic in the variable $\phi$ 
and $j_z$ is restricted to $-1\le j_z\le1$. In these figures we can see two 
different dynamical regimes: closed and open orbits separated by a separatrix 
corresponding to the classical energies for $j_z=\pm1$. The closed orbits 
correspond to classical energies less than $h(j_z=-1)$ or greater than 
$h(j_z=1)$. The arising of the separatrix in phase space is associated to the 
level approximation in the quantum spectrum \cite{reis,gian}. This can be seen 
in the three cases below:\\

\noindent
{\em (i)} $g'/\delta'=0$ ($g/\delta=0$): We observe the appearance of two 
separatrices for $G'/\delta'>0.125$ ($G/\delta>0.001$) next to upper 
and lower classical energies $h(j_z=1)$ and $h(j_z=-1)$, while in the quantum 
spectrum we observe the level approximation in both extreme energies.\\

\noindent
{\em (ii)} $g'/\delta'= 0.375$ ($g/\delta=0.003$): We observe the 
appearance of a separatrix for $G'/\delta'>0.125$ ($G/\delta>0.001$) 
next to the lower classical energy $h(j_z=-1)$ and for $G'/\delta'>0.3125$ 
($G/\delta>0.0025$) next to the upper classical energy $h(j_z=1)$, 
while in the quantum spectrum we observe respective level approximations next 
to 
upper and lower energies.\\

\noindent
{\em (iii)} $g'/\delta'= -0.375$ ($g/\delta= -0.003$):  Besides the 
behavior seen in {\em (ii)}, we also observe a degeneracy of two classical 
trajectories with the same energy. This degeneracy disappears for 
$G'/\delta'>0.0625$ ($G/\delta>0.0005$).\\

The previous aspects can be explained analytically by means of the critical 
points of the function $h(\phi,j_z)$. These are of two kinds:

\noindent
{---}Maxima or minima localized at \begin{equation}
(\phi,j_z)_{\mathrm{max,min}}=\left(2n\pi,\frac{\delta'+2g'}{8G'-2g'}\right) 
\label{max_min1}
\end{equation}
\noindent
and \begin{equation}
(\phi,j_z)_{\mathrm{max,min}}=\left((2n+1)\pi,-\frac{\delta'+2g'}{8G'+2g'}\right), 
\label{max_min2}
\end{equation}

\noindent
{---}Saddle points at \begin{equation}
(\phi,j_z)_{\mathrm{saddle}}=\left(\arccos\left[\frac{\delta'+4g'}{8G'}\right],1\right)
\end{equation} and \begin{equation} 
(\phi,j_z)_{\mathrm{saddle}}=\left(\arccos\left[\frac{-\delta'}{8G'}\right],
-1\right).\end{equation}

These critical points are shown in Figs. \ref{hclass1}--\ref{hclass4}. The 
critical points in Eqs. \eqref{max_min1} and \eqref{max_min2} can be maxima or 
minima, depending on whether $g'/\delta'\geq0$ or $g'/\delta'<0$. The closed 
orbits are created around the  maximum and minimum points. Therefore, the 
conditions for the existence of these points are the conditions for the arising 
of the separatrix for upper and lower classical energies. We deduce from the 
conditions above that we have the arising of the two separatrices starting for 
the same value of the three-mode coupling $G'/\delta'=0.125$ when 
$g'/\delta'=0$. In the same way, when $g'/\delta'=0.375$ we have two distincts 
values of the three-mode coupling $G'/\delta'=0.125$ and $G'/\delta'=0.3125$ 
for the arising of the separatrix for upper and lower classical energies 
respectively. We can interpret these values as the critical values of the 
quantum phase transitions, with the appearance of the closed 
orbits in phase space signaling the possibility of another physical phase 
accessible to the system. As it is shown in Figs. \ref{hclass1}--\ref{hclass4},
 the arising of this second phase for upper classical energies 
depends on the value of the parameter $g'/\delta'$. For $g'/\delta'=0$, the 
second phase (closed orbits) arises for upper energies when 
$G'/\delta'\ge0.125$, while for $g'/\delta'=0.375$ and $-0.375$ it occurs for 
$G'/\delta'\ge 0.3125$ and $G'/\delta'\ge 0$, respectively. 

\section{Quantum phase transition}

\begin{figure}\centering
 \includegraphics[width=8.5cm,angle=0]{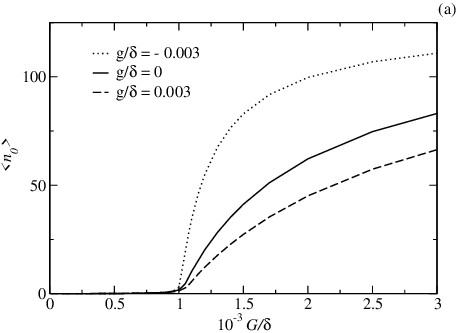}\vspace{.55cm}
 \includegraphics[width=8cm,angle=0]{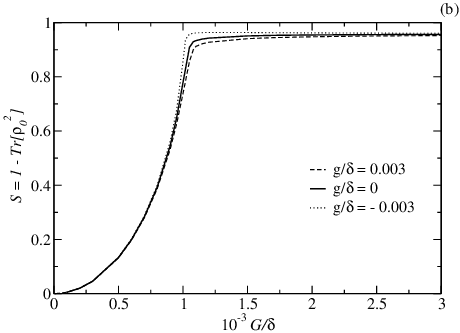}
 \caption{Average value of population $\langle n_0\rangle$ (a) and the linear 
entropy $S=1-\mathrm{Tr}[\rho_{n_0}^2]$ of  the $0$-mode (b) as a function of the 
three-mode coupling $G/\delta$ for the ground-state of the system.}
 \label{transition}
\end{figure}

We can observe characteristics of a quantum phase transition by looking at the 
properties of the ground state and of the measures of its entanglement 
\cite{QPT,ordem}. In Fig. \ref{transition} we show the average value of the 
$0$-mode population $\langle n_0\rangle$ and the linear entropy 
$S=1-Tr[\rho_{n_0}^2]$ for the ground state of the system. Here, $\rho_{n_0}$ 
stands for the reduced density operator of the $0$-mode for the ground state, 
obtained by tracing out the other modes. We clearly observe two distinct 
behaviors for these quantities, for $G/\delta<0.001$ or 
$G/\delta>0.001$. This can be associated with the phase transition 
between the two dynamical regimes of macroscopic self-trapping (MST) and {of 
oscillations} (RO). For $G/\delta<0.001$, the average $0$-mode 
population is practically null and we can deduce that the total population of 
bosons is on average in the $1$-mode. This situation characterizes the MST 
phase in which the bosons remain in a single mode. This more organized phase 
has 
a relatively small linear entropy, so a small entanglement. For 
$G/\delta>0.001$, the average $0$-mode population is non-null and 
increases with the three-mode coupling. In this situation we can say that the 
total population of bosons tends on average to be divided among the different 
modes. This situation characterizes the RO phase in which the bosons do not 
have a preferred mode and oscillate between them. As the system is in a more 
disordered situation the linear entropy is larger, and so is the entanglement.

\subsection{Order of the phase transition}

\begin{figure}\centering
 \includegraphics[width=8.5cm,angle=0]{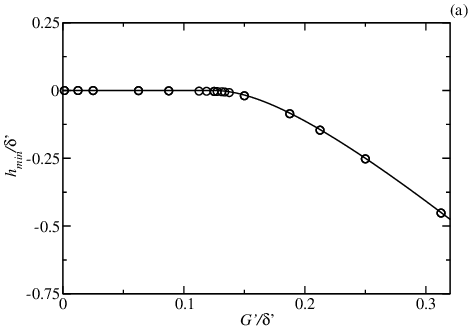}\vspace{.55cm}
 \includegraphics[width=8.5cm,angle=0]{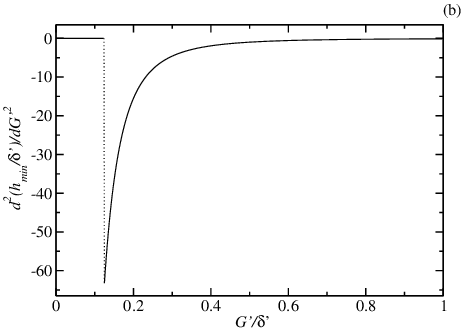}
 \caption{The ground state energy (a) and its second derivative (b) as a 
function of the three-mode coupling $G'/\delta'$. In (a) we also observe the 
ground state energy rescaled as $E/(N/4)$ (circle).}\label{energia}
\end{figure}

\begin{figure}\centering
 \includegraphics[width=8cm,angle=0]{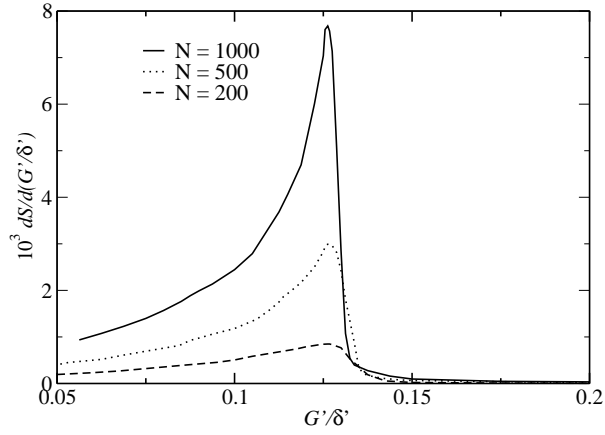}
 \caption{The first derivative of the linear entropy $S=1-Tr[\rho_{n_0}^2]$ as 
a 
function of the three-mode coupling $G'/\delta'$ for the ground state and 
different total boson populations $N=200$, $500$, and $1000$.}\label{dentropia}
\end{figure}

We can now discuss the order of the quantum phase transition. Following the 
usual criterion \cite{QPT,ordem}, we can obtain this information by analyzing 
the behavior of the ground state energy and the entanglement, and their 
derivatives. Figure \ref{energia} shows the behavior of the ground state energy 
as a function of the three-mode coupling $G'/\delta'$. In the same figure we 
observe the minimum of the corresponding classical energy $h(j_z,\phi)$, which 
shows a perfect agreement with the ground state energy rescaled as $E/(N/4)$. 
Taking the first and second derivatives, we observe that the first derivative 
is 
continuous and the second one is discontinuous [Fig. \ref{energia} (b)]. In 
Fig. 
\ref{dentropia} the first derivative of the linear entropy as a function of the 
three-mode coupling $G'/\delta'$ is shown for different values of the total 
boson population $N$. We observe that the curves tend to diverge as 
$N\rightarrow\infty$ at the critical value $G'/\delta'=1.25$. Such aspects, 
i.e., the second derivative of the energy and the first derivative of the 
entropy, both discontinuous for the ground state, characterize a second order 
phase transition. We note that a first-order transition would be characterized 
by a discontinuous first derivative of the ground state energy \cite{QPT,ordem}.

\subsection{Phase diagram}

\begin{figure}\centering
  \includegraphics[width=8cm,angle=0]{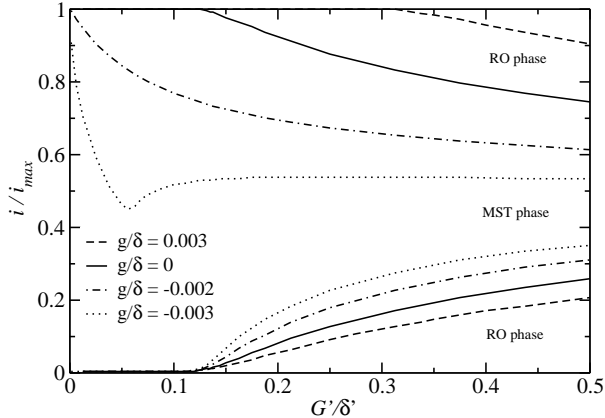}
  \caption{Phase diagram for $g/\delta=0.003,\;0,\;-0.002$,   
and $-0.003$. For each value of $g/\delta$ we have two lines: in the 
middle of the lines the eigenstates are in a phase of macroscopically 
self-trapping (MST), above and below these lines the eigenstates are in a phase 
{of the regime of oscillations} (RO).}\label{diagramaB}
\end{figure}

We can now construct a phase diagram for the transition. We observe that in the 
quantum spectrum (Fig. \ref{espectro}) for $G/\delta<0.001$ the 
eigenenergies are restricted to values $0<E<N/2+(g/\delta)(N/2)^2$, while for 
$G/\delta>0.001$ some eigenenergies $E<0$ and 
$E>N/2+(g/\delta)(N/2)^2$ arise. These quantum states correspond precisely to 
classical closed orbits which arise for $G/\delta>0.001$. In this way 
we can associate these quantum states with the second phase of the system and 
separate the spectrum in different regions for the values $E<0$ and 
$E>N/2+(g/\delta)(N/2)^2$. Varying the three-mode coupling $G/\delta$, we 
change 
the number of eigenstates (regions of the spectrum) associated with the second 
phase. While $G/\delta$ increases, the second phase region of the spectrum 
becomes larger. This phase diagram is shown in Fig. \ref{diagramaB}, where the 
vertical axis is a relative index $i/i_{\mathrm{max}}$ of the eigenvalues $E_i$. The 
lower curve separates the eigenvalues $E_i<0$ while the upper curve separates 
the eigenvalues $E_i>N/2+(g/\delta)(N/2)^2$. We can observe that for 
$g/\delta\neq0$ the arising of the second phase in upper energies is changed. A 
similar phase diagram had already been done for the Lipkin model in Ref. 
\cite{lipkin}. The difference between our three-mode model and the Lipkin model 
is just the asymmetry due to coefficient $g/\delta$, since the Lipkin model has 
a symmetric phase diagram. {This feature gives rise to a situation in this 
three-mode model which does not occur in the Lipkin model: the manifestation of 
the second phase in the upper energies is changed for $g/\delta\neq0$}.

\section{Application}

In the previous sections we have analyzed the behavior of the three-mode 
Hamiltonian and of its corresponding classical hamiltonian function for 
different values of the parameters $G/\delta$ and $g/\delta$. But until now we 
have not discussed how we can vary physically these parameters. We 
will examine below an example for exciton-polaritons in a semiconductor 
microcavity.

\begin{figure}\centering
  \includegraphics[width=8cm,angle=0]{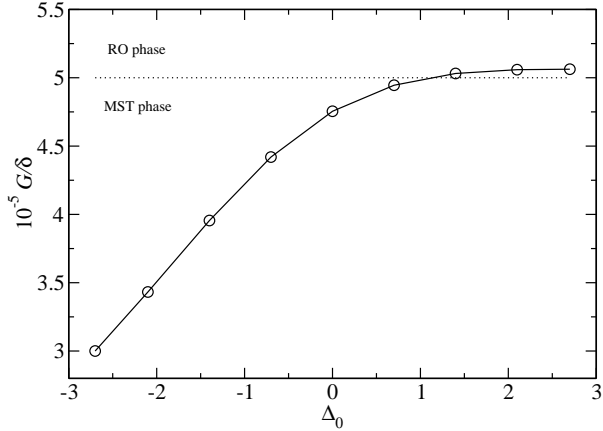}
  \caption{Numerical value for $G/\delta$ as a function of the detuning 
$\Delta_0$, calculated from experimental parameters of a semiconductor 
microcavity \cite{Eduardo}. The horizontal dotted line marks the critical value 
$G/\delta=0.00005$ for $N=10^4$.}\label{polariton}
\end{figure}

We can use experimental values of an exciton-polariton system at the magic 
angle configuration to test the results obtained in the previous sections. The 
values of the parameters $G/\delta$ and $g/\delta$ in the effective Hamiltonian 
Eq. \eqref{H2} can be determined from the parameters $\hbar\Omega_k$ and 
$V_{\mathbf{k,k',q}}^{PP}$ in the exciton-polariton Hamiltonian Eq. \eqref{Hlow}. 
In Fig. \ref{polariton} we observe a curve of $G/\delta$ as a function of the 
detuning $\Delta_0$ numerically calculated from experimental values 
\cite{Eduardo}. In this figure we note that the increase in the three-mode 
coupling due to detuning variation results in phase transition from MST to RO. 

We emphasize that an exciton-polariton system is a very complex system with 
numerous features which were not considered in our simple analysis. For 
example, the coexistence of other system components such as photons, excitons, 
or bi-excitons, and the consequent conversion between them. Also, there is the 
non-conservation of total number of particles due to pumping and dissipation in 
the system. In this way, our work constitutes a first and simple approach to a 
complex system and is based on features of the macroscopic population of the 
three exciton-polariton modes at the magic angle configuration. In other words, 
we have an effective but physically meaningful model.

\section{Conclusions}

The three-mode Hamiltonian studied in this work can be classified in a class of 
schematic models named Curie-Weiss models. There are many examples of this 
class 
as the Lipkin model and the pairing model (both in nuclear physics), the Dicke 
model for the superradiance, and others. Models in this class share the  
characteristic of a Hamiltonian which allows an expansion in powers of $1/N$, 
which in a thermodynamic limit leads to a classical Hamiltonian analysis. In 
these models a quantum phase transition is signaled by a level approximation 
near an inflection point of the spectrum. This feature in the classical 
analysis 
is signaled by the appearance of a separatrix in the phase space. In the 
Curie-Weiss class, the Lipkin model has the particularity of a symmetric level 
approximation in both upper and lower inflection points of the spectrum. The 
three-mode Hamiltonian studied here also has this particularity and allows 
moreover
for a situation ($g/\delta\neq0$) of an asymmetric level approximation in the 
upper and lower parts of the spectrum. In other words, we can have an asymmetric 
phase diagram. 

The effective Hamiltonian obtained in this work can describe exciton-polaritons 
in a semiconductor microcavity at the magic angle. In terms of the 
exciton-polariton system, our results constitute an effective but physically 
meaningful model. It is a simple approach to this intricate system, based on 
the 
macroscopic features of the exciton-polariton population of only three modes.

\begin{acknowledgments}

We would like to thank CNPq (Grant No. 312207/2015-8) and CAPES for financial
support. H.M.F., J.G.P.F. and G.Q.P. would like to dedicate this work to the
memory of Maria Carolina Nemes.

\end{acknowledgments}

\end{document}